\documentclass[useAMS,usenatbib]{mn2e}

\usepackage{graphicx}

\newcommand{\dxdy}[2]{{\frac{\partial{#1}}{\partial{#2}}}}

\newcommand{\DxDy}

\title[Gravity Waves in the Sun]{Gravity waves in the Sun}
\author[Tamara M. Rogers and Gary A. Glatzmaier]{Tamara M. Rogers$^{1}$\thanks{E-mail:trogers@pmc.ucsc.edu} and Gary A. Glatzmaier$^{2}$\\
$^{1}$Astronomy and Astrophysics Department, University of California, Santa Cruz, 95060 CA USA\\
$^{2}$Earth Sciences Department, University of California, Santa Cruz, CA 95060 USA}
\begin{document}

\pagerange{\pageref{firstpage}--\pageref{lastpage}} \pubyear{2002}

\maketitle

\label{firstpage}

\begin{abstract}
We present numerical simulations of penetrative convection and gravity wave excitation in the Sun.  Gravity waves are self-consistently generated by a convective zone overlying a radiative interior.  We produce power spectra for gravity waves in the radiative region as well as estimates for the energy flux of gravity waves below the convection zone.  We calculate a peak energy flux in waves below the convection zone to be three orders of magnitude smaller than previous estimates for m=1.  The simulations show that the linear dispersion relation is a good approximation only deep below the convective-radiative boundary.  Both low frequency propagating gravity waves as well as higher frequency standing modes are generated; although we find that convection does not continually drive the standing g-mode frequencies.    
\end{abstract}


\section{Introduction}
Convection bounded by a stable region is important in all aspects of stellar evolution.  The depth to which convective motions can overshoot into an adjacent stable layer has important consequences for many stages of stellar evolution, from dredge up in late stages of stellar evolution \citep{kip94} to nucleosynthesis in Classical Novae \citep{woo86,ke98,ke99,al04}.  In the Sun, the overshoot region at least partially overlaps the tachocline (the shear layer at the base of the convection zone where the rotation changes from being differential in the convection zone to solid body in the radiation zone and the likely site of toroidal field generation in the dynamo process).  Therefore, understanding the extent of the overshoot is necessary for understanding the nature of the tachocline; is it a violent region, constantly bombarded by overshooting plumes, or is it a laminar region just below the overshoot?  The answer to this question can constrain models for the solid body rotation of the solar interior.

In addition to the uncertainty of penetrative convection, which has plagued stellar evolution theory, there are other noteworthy aspects of convective-radiative boundaries.  Convection adjacent to a radiative region generates internal gravity waves which can have  dynamical affects.  In the Earth's atmosphere, the nonlinear interaction between convection, shear and gravity waves is the cause of the quasi-biennial oscillation (QBO) \citep{bal01}; this shear-gravity wave interaction has been demonstrated in a remarkable experiment by \cite{pmb78}.  The same process responsible for the QBO has been invoked to explain the solar magnetic cycle \citep{ktz99}.  

Internal gravity waves can share angular momentum with the mean flow when they are attenuated.    The transport and deposition of angular momentum by gravity waves is one explanation for the solid body rotation observed in the radiative interior of the Sun \citep{kq97,ztm97}.  However, a vast literature in atmospheric science tells us that gravity waves have an anti-diffusive nature; they enhance local shear rather than smooth it.  In addition, internal gravity waves have also been implicated in the mixing of species \citep{pre81}, and in order to explain the lithium depletion \citep{sch93,gls91}.

Most studies of convection bounded by a stable layer have been focused on the penetration depth.  The depth that convection overshoots into an adjacent radiative region has been studied extensively by numerical simulations in two \citep{mas84,htm86,htmz94,rg05} and three dimensions \citep{bct02,ssh00}.  In general, these simulations find that the penetration depth is a sensitive function of the stiffness of the subadiabatic layer and find that it is also sensitive to the Peclet number ($\frac{UL}{\kappa}$) \citep{bct02}.  Early 2D studies with ``soft'' interfaces (mild subadiabaticity) \citep{htm86,htmz94} found that overshooting convection could extend the adiabatic region.  Three-dimensional simulations \citep{bct02} and simulations with very stiff (large subadiabaticity) radiative regions \citep{rg05} find no extended adiabatic region, in agreement with helioseismic observations.  However, in all of the previous studies a constant (with height) subadiabaticity was used and none were as stiff as the Sun.  Analytic models generally predict extended adiabatic regions, but rely on highly uncertain parametrizations.  The penetration depth in the Sun and its dependencies will be discussed in greater detail in a forthcoming paper.  

While internal gravity waves have been studied analytically for some time, e.g. \cite{pre81}, they have not received much attention in numerical simulations of stellar interiors.  In the few instances where they have been simulated they have not been analyzed in detail and have been unlike those in the Sun because the subadiabaticity was much less than solar.  However, \cite{kir03} did extensive numerical studies in 2D, similar to those by \cite{htmz94} and analyzed the wave spectrum produced and compared it to previous analytic predictions for g-mode spectra by \cite{gls91}.  They found that the spectrum of waves produced in numerical simulations was more broadband in frequency and that numerical simulations always predicted larger fluxes in waves than analytic studies.  The overestimate of energy in waves in that study was probably due to the mild subadiabaticity used in those calculations.  The subadiabaticity of the stable region is extremely important in these types of simulations \citep{rg05}; the mild values used in all previous simulations allow plumes to penetrate much more deeply, causing larger nonlinearities and velocities than would otherwise occur.

However, since internal gravity waves may be maintaining the solid body rotation of the solar interior \citep{ztm97}, may be driving a QBO like oscillation (coined SLO by \cite{tc05} ) at the base of the convection zone \citep{ktz99} and may someday be observed with helioseismology \citep{afa00}, we have begun a detailed study via numerical simulations of their excitation and evolution.  The way in which gravity waves redistribute angular momentum and the likelihood of their observation depends on the spectrum of gravity waves produced by the overlying convection, their nonlinear interaction with convection and rotation, and on their radiative damping.  Here we discuss the spectrum of gravity waves produced by overlying convection and penetration, radiative damping and the energy flux of waves beneath the convection zone.  The impact gravity waves have on the angular velocity in the radiative interior will be discussed in a forthcoming paper.  We discuss the numerical method in section 2, convection in section 3 and gravity waves in section 4. 

\section[]{Numerical Model}
\subsection{Equations}

We solve the Navier Stokes equations with rotation in the anelastic approximation \citep{go69} for an ideal gas.  The anelastic approximation is appropriate when the fluid flow is sufficiently subsonic and the perturbations relative to the mean (reference) thermodynamic state are small.  The radially (r) dependent reference state is taken from a 1D solar model (Christensen-Dalsgaard private communication).  Unlike previous formulations of the anelastic equations \citep{ga85,rg03}, here our reference state temperature gradient is not adiabatic, and we choose to use the temperature perturbation as our working thermodynamic variable, instead of the specific entropy.  This formulation has advantages and disadvantages compared to the standard formulation \citep{rg05}.  The equations are solved in 2D cylindrical coordinates, with the radiation zone extending from 0.001R$_{\odot}$ to 0.718R$_{\odot}$ and the convection zone occupying the region from 0.718R$_{\odot}$ to 0.9R$_{\odot}$.  This geometry represents an equatorial slice of the Sun that includes most of the radiative and convective regions.  For numerical reasons we put another stable region from 0.9R$_{\odot}$ to 0.93R$_{\odot}$.  The exclusion of $\frac{1}{10}\%$ of $R_{\odot}$ around the centre has little effect since gravity waves internally reflect where their frequency equals the Brunt-Vaisala frequency, which vanishes at the centre.  

We solve the following anelastic equations for perturbations to the reference state: 

\begin{equation}
\nabla \cdot \overline{\rho} \vec{v} = 0 .
\end{equation}

\begin{eqnarray}
\lefteqn{\dxdy{\vec{v}}{t}+(\vec{v}\cdot\nabla)\vec{v}=-\nabla P - Cg\hat{r} + 2(\vec{v}\times\Omega)+}\nonumber\\
&    &\overline\nu(\nabla^{2}\vec{v}+\frac{1}{3}\nabla(\nabla\cdot\vec{v}))
\end{eqnarray}

\begin{eqnarray}
\lefteqn{\dxdy{T}{t}+(\vec{v}\cdot\nabla){T}=-v_{r}(\dxdy{\overline{T}}{r}-(\gamma-1)\overline{T}h_{\rho})+}\nonumber\\
&  & {(\gamma-1)Th_{\rho}v_{r}+\gamma\overline{\kappa}[\nabla^{2}T+(h_{\rho}+h_{\kappa})\dxdy{T}{r}]+}\nonumber\\
&  & \gamma\overline{\kappa}[\nabla^{2}\overline{T}+(h_{\rho}+h_{\kappa})\dxdy{\overline{T}}{r}] + \frac{\overline{Q}}{c_{v}}
\end{eqnarray}

Equation (1) represents the continuity equation in the anelastic approximation, where $\overline{\rho}$ is the reference state density and $\vec{v}$ is the fluid velocity.  Equation (2) is the momentum equation, assuming a constant dynamic viscosity, {$\rho\nu$}.  The gravitational acceleration, -$\overline{g}\hat{r}$ is taken from the 1D solar model.  The rotation rate, $\Omega$ is set to the mean rotation rate of the Sun $2.6\times10^{-6}$ rad/s.  The reduced pressure, P \citep{bra95} is defined as 
 
\begin{equation}
 P=\frac{\it{p}}{\overline{\rho}} +U
\end{equation} 
where $\it{p}$ is the pressure perturbation and U is the gravitational potential perturbation. The co-density perturbation is defined as: 

\begin{equation}
C=-\frac{1}{\overline{T}}(T+\frac{1}{g\overline{\rho}}\frac{d\overline{T}}{dz}\it{p}).
\end{equation}

Equation (3) is derived from
\begin{equation} 
\frac{dT}{dt}=(\dxdy{T}{\rho})_s\frac{d\rho}{dt} + (\dxdy{T}{S})_{\rho}\frac{dS}{dt}
             = -\frac{\gamma-1}{\alpha}\vec{\nabla}\cdot\vec{v} +\frac{1}{\rho c_{v}}\vec{\nabla}\cdot(c_{p}\rho\kappa\vec{\nabla}T)
\end{equation}
where here the thermodynamic variables are the full variables (sum of the reference state, denoted by an overbar, and the perturbation) and an ideal gas is assumed.  The thermal expansion coefficient $\overline{\alpha} = \frac{1}{\overline{T}}$, the adiabatic exponent $\overline{\gamma}=\frac{c_{p}}{c_{v}}$ and $\overline{\kappa}$, the thermal diffusivity are all functions of radius.  The inverse density scale height and thermal diffusivity scale height are defined as:

\begin{equation} 
h_{\rho}=\frac{1}{\overline{\rho}}\frac{d\overline{\rho}}{dr} \hspace{1cm} and \hspace{1cm} h_{\kappa}=\frac{1}{\overline{\kappa}}\frac{d\overline{\kappa}}{dr}
\end{equation}

The first term on the right hand side of Equation (3) can also be written as 
\begin{equation}
-v_{r}h_{\rho}\overline{T}(\dxdy{ln\overline{T}}{ln\overline{\rho}}-(\dxdy{lnT}{ln\rho})_{ad})
\end{equation}
which illustrates how this term depends on the local difference between the nonadiabatic reference state and an adiabatic state.  The last two terms in Equation (3), which are only a function of the reference state, are assumed zero at the beginning of our simulations.  We neglect viscous heating for these low viscosity simulations.  The top and bottom boundary conditions are isothermal, stress-free and impermeable.

\subsection{Numerical Technique}

The technique used here is similar to that of \cite{rg05}, but for a rotating disk instead of a box.  For numerical simplicity we take the curl of the momentum equation, leaving us with a vorticity equation:

\begin{equation}
\dxdy{\omega}{t}+(\vec{v}\cdot\vec{\nabla})\omega=(2\Omega + \omega)h_{\rho}v_{r}-\frac{g}{\overline{T}r}\dxdy{T}{\theta}-\frac{1}{\overline{\rho}\overline{T}r}\dxdy{\overline{T}}{r}\dxdy{{\it{p}}}{\theta} +\overline{\nu}\nabla^{2}\omega
\end{equation}

where the vorticity, $\vec{\omega}=\vec{\nabla}\times\vec{v}$ is in the $\hat{z}$ direction.  We have neglected viscous terms involving $\dxdy{\overline{\nu}}{r}$.  We calculate the pressure term using the longitudinal component of the momentum equation (2):

\begin{equation}
\frac{1}{\overline{\rho}r}\dxdy{\it{p}}{\theta}=-\dxdy{v_{\theta}}{t}-(\vec{v}\cdot\vec{\nabla}\vec{v})_{\theta}+\overline{\nu}[(\nabla^{2}\vec{v})_{\theta} - \frac{h_\rho}{3r}\dxdy{v_{r}}{\theta}] .
\end{equation}
where $v_{\theta}$ and $v_{r}$ represent the longitudinal and radial velocity components, respectively.  The time derivative of $v_{\theta}$ is from the previous timestep and we neglect the perturbation to the gravitational potential in (10).  

These equations are solved using a Fourier spectral transform in the longitudinal ($\theta$) direction and a finite difference scheme on a non-uniform grid in the radial (r) direction.  Time advancing is done using the explicit Adams-Bashforth method for the nonlinear terms and an implicit Crank-Nicolson scheme for the linear terms.  

We do a 21st order polynomial fit to the reference state density ($\overline{\rho}$), temperature ($\overline{T}$), mass (which gives us gravity, $\overline{g}$), opacity ($\overline{k}$) and $\overline{\gamma}$ given in the 1D solar model.  Figure 1 shows density and temperature as a function of radius in our model (solid line) compared to the standard model (dotted line).  Analytic derivatives of these (fitted) values give us inverse scale heights ($h_{\rho}, h_{T}$ and $h_{\kappa}$) and the temperature gradient.  

The subadiabaticity in the radiative region is then defined as:
\begin{equation}
\dxdy{\overline{T}}{r} - (\overline{\gamma} -1)h_{\rho}\overline{T}
\end{equation}
where the second term in equation (11) is the adiabatic temperature gradient.  Figure 2 shows the effective Brunt-Vaisala frequency (subadiabaticity), defined in (16) as a function of radius.  The superadiabaticity in the convection zone is specified to be a constant, $10^{-7}$.  The thermal diffusivity is that given by the solar model, multiplied by a constant factor for numerical stability:
 
\begin{equation}
\overline{\kappa}=kapmult*\frac{16\sigma \overline{T}^{3}}{3\overline{\rho}^{2}\overline{k}c_{p}}
\end{equation}
where $\sigma$ is the Stefan-Boltzman constant and the multiplying factor, $\it{kapmult}$, is $10^{5}$ (and therefore, the convective flux is $10^{5}$ larger than Solar).  This gives us the proper radial profile of the radiative diffusivity, albeit increased by a large factor for numerical stability; this is a ``turbulent'' diffusivity.  The viscous diffusivity is specified to be:
\begin{equation}
  \overline{\nu}=\frac{const}{\overline{\rho}}
\end{equation}
keeping the dynamic viscosity constant.  Since both $\overline{\nu}$ and $\overline{\kappa}$ vary with height, the Pr (Prandtl number = $\frac{\overline{\nu}}{\overline{\kappa}}$) varies with height from $10^{-2}_{BCZ}$ (at base of the convection zone) to $0.7_{TCZ}$ (at top of convection zone) and is $10^{-3}$ near the centre.  Therefore, the Ra (Rayleigh number = $\frac{\overline{g} \alpha\Delta\nabla T D^{4}}{\overline{\nu}\overline{\kappa}}$, where D is the depth of the convection zone and $\Delta\nabla T$ is the superadiabatic temperature gradient) also varies from $\approx 10^{8}$ at the base of the convection zone to $\approx 10^{7}$ at the top of the convection zone. The Ekman number (Ek=$\frac{\overline{\nu}}{2\Omega D^{2}}$) varies from $10^{-4}_{BCZ}$ to $10^{-2}_{TCZ}$.  The resolution used in these models is 2048 longitudinal zones x 1500 radial zones, with 500 radial zones dedicated to the radiative region.  

\section[]{Convection Bounded by a Stable layer}

Here we discuss the convection in this model.  In order to illustrate the differences between convection bounded by a radiative region and convection bounded by a hard boundary we also present a model that has an impenetrable bottom boundary.  The model with an impenetrable boundary has the same reference state as the model with a stable region and therefore, has the same Ra, Pr and Ek.  We note here that the simulations with impenetrable boundaries usually require greater spatial resolution near the boundaries and smaller timesteps because the hard boundaries deflect fast descending (ascending) plumes into shallow, high-speed horizontal flows.  This problem is more severe at the top of the computational domain because the density is lower there, resulting in larger velocities.  This is one reason we have added a very shallow stably stratified region at the top boundary.  Another is that it is more realistic to have rising plumes stopped by a stable layer rather than an artificial hard boundary.

Snapshots of the vorticity in the two models are shown in Figure 3.  There are many differences immediately obvious in this figure.  Most noticeable is the nature of the convective cells.  In the model with an impenetrable (hard) bottom boundary, there are fewer convective cells and those cells are more orderly and appear more laminar.  Because descending and ascending plumes are diverted horizontally by the hard, stress-free boundary nearly all of their vertical kinetic energy is converted into horizontal kinetic energy.  These (relatively) large horizontal velocities allow the flow to travel further before cooling (heating) enough to fall (rise).  In the model bounded below by a stable layer, a small fraction of the vertical energy in descending plumes is transferred to the stable region generating gravity waves.  Because of this transfer of energy and less rigid horizontal trajectories, horizontally diverted fluid does not travel as far before cooling (heating) and sinking (rising).  The time dependent convective penetration in the model bounded by a stable region also results in more chaotic interaction between convective cells, albeit with lower overall kinetic energy.  

Figure 4 shows the kinetic energy density ($v_{\theta}^{2}+v_{r}^{2}$), horizontal kinetic energy flux $v_{\theta}*(v_{\theta}^{2}+v_{r}^{2})$ and vertical kinetic energy flux $v_{r}*(v_{\theta}^{2}+v_{r}^{2})$, all scaled to the maximum values in the impenetrable case. It is obvious in this figure that in the convective region there is significantly more energy (by 2-3 orders of magnitude) in the model with hard boundaries.  This is due almost entirely to a difference in magnitude of horizontal velocity (4b, note magnitudes), as explained above.  

In models bounded by a stable region a fraction of the kinetic energy is shared with the radiative region.  Descending plumes overshoot the convective-radiative boundary and energy is transferred to gravity waves.  The simplest estimate of the penetration depth is given where the kinetic energy density drops to 1\% of its maximum value in the convection zone.  In this model this results in a penetration depth of approximately 0.1 pressure scale heights, but no extended adiabatic region.  Several things can affect this penetration depth.  First, the ``stiffness'' of the subadiabatic layer has been shown to be of utmost importance in several previous numerical studies \citep{htm86,bct02,rg05}.  As stated above, this model has the (realistic) subadiabatic structure of the Sun and hence, a very high ``stiffness'', which makes the penetration depth small.  The dimensionality of the model also has an effect; 3D models have smaller penetration depths than 2D models \citep{bct02}.  Therefore our estimate may be an upper limit.  In addition, rotation has been shown to hinder penetration \citep{bct02,jlww96}, while the effect of the Rayleigh number is still unclear \citep{rg05} and may depend on the dimension.  

Kinetic energy spectra in the convection zones of the two models also differ.  Figure 5 shows the kinetic energy as a function of horizontal wavenumber at a radius half way through the convection zone in both models.  As illustrated in Figure 4, and now again in 5, the overall kinetic energy in the convection zone is larger for the model with an impenetrable lower boundary (shown as the dotted line in Figure 5), especially at low wavemode (large scales).  Both models are fit rather well with a $k^{-4}$ power law.  While $k^{-3}$ spectrum is expected in 2D turbulence, the geometry of the flow has not been previously considered, nor the effect of adjacent stable regions (both models have stable regions on top).  Fully convective simulations of convective motion in a 2D disk give a variety of scaling laws (between $k^{-2}$ and $k^{-4}$) depending on the Ekman number, the density stratification and the depth at which the spectrum is calculated.  The model with hard boundaries shows less of a dissipation range and some build-up at small scales.  The model bounded by a stable region shows a dissipation range and no such build-up at large wavemodes.  This difference is probably due to the issues mentioned above: vorticity generation and large horizontal velocities along the (impenetrable) boundary.  
   
In summary, both the energetics and the character of convection are changed when a bounding stable layer is used instead of the traditional (artificial) impenetrable boundary.  

\section[]{Gravity Waves}

Gravity waves can be generated by Reynolds stresses in a convective region overlying a radiative region as well as by descending plumes that overshoot the convective-radiative boundary.  The gravity waves generated by such processes have been studied previously in analytic studies.  In particular, \cite{gmk94} studied the spectrum of gravity waves generated by Reynolds stresses in the overlying convective region and \cite{gls91} studied the spectrum of gravity waves generated by overshooting plumes.  Both models are analytic and assume some combination of mixing length theory (MLT) and a Kolmogorov spectrum for turbulence.  

Gravity waves have the ability to mix species and transport angular momentum and thus, may have profound effects on the dynamics of the radiative zone in the Sun.  In the following, we discuss results from both linear and nonlinear numerical simulations.  First, however we briefly review basic properties of gravity waves in the asymptotic regime without rotation (for a more thorough review see, e.g. \cite{un89,cd02,go91}).  The asymptotic, linear dispersion relation for pure gravity waves is: 

\begin{equation} 
\omega=\frac{Nk_{h}}{\hat{k}}
\end{equation} 
where $k_{h}$ represents the horizontal wavenumber and 
\begin{equation}
\hat{k}=(k_{r}^2 + k_{h}^2)^{\frac{1}{2}}
\end{equation} 
where $k_{r}$ is the radial wavenumber.  The Brunt-Vaisala frequency, N, represents the upper limit on the frequency of gravity waves.  It depends on the thermal stratification of the fluid and is defined as: 
\begin{equation}
N^{2}=\frac{\overline{g}}{\overline{T}}(\dxdy{\overline{T}}{r} - (\overline{\gamma} -1)h_{\rho}\overline{T})  
\end{equation}
From the basic linear dispersion relation (14) it is seen that the frequencies of gravity waves depend only on the gas stratification and the direction of phase propagation.  From the above dispersion relation it can also be shown that the group velocity and phase velocity are perpendicular, a peculiar trait of internal gravity waves.  
 
In addition to the standing g-modes (in the frequency range of $100\mu$Hz to $500\mu$Hz), propagating waves from the overlying convection and penetration are produced.  These propagating gravity waves, generally with much lower frequencies, have been suggested as the source of the oscillating shear layer at the base of the convection zone, the net angular momentum transport in the radiative interior \citep{ztm97,ktz99,tkz02} and the mixing of species \citep{gls91}.  The spectrum of these waves has been predicted in those analytic studies.  However, here we present the first self-consistent nonlinear simulation and analysis of these waves in the Sun's radiative interior.

\subsection{Linear Gravity Waves}

While our main simulation here is nonlinear in both the convective and radiative regions of the computational domain, we have computed linear cases for comparison with previous asymptotic results and to illustrate the differences between linear and nonlinear solutions.  We first discuss the linear calculations. Figure 6 shows a snapshot of the vorticity in the radiation zone produced with a linear version of our model, excluding the convection zone and initially perturbed at several horizontal wavenumbers.  While there are several different horizontal wavenumbers depicted in this image, their modes do not interact.  This is similar to that predicted from ray-tracing \citep{go91}.  

In our partially linear simulation, the full set of nonlinear equations is solved in the convection zone, but the nonlinear terms in equations (3), (9) and (10) are decreased linearly between 0.718$R_{\odot}$ (the radius of the transition between convective and radiative layers) and 0.69$R_{\odot}$.  Therefore, below 0.69$R_{\odot}$ only the linear set of equations is solved.  A snapshot of vorticity looks similar to that in 8a (the fully nonlinear case).  However, there is a remarkable difference between this gravity wave pattern (seen in 8b) and that in figure 6.  When constantly being driven by penetrative convection at many different frequencies and wavenumbers g-modes form a much simpler spiral pattern in the radiative interior.  

A simple reason for this behaviour can be explained as followed.  If convection is considered as nearly equally spaced convective cells, descending plumes drive similar wavenumbers, albeit slightly out of phase.  The nonlinear wave-wave interactions generate new waves with wavenumbers equal to both the sum and difference of the original waves.  The high wavenumber waves are damped in a very short distance; wherease the small wavenumber (long wavelength) waves survive through most of the radiative interior.  This wave-wave interaction only occurs in nonlinear simulations.  Figure 6 has no nonlinear interactions and thus all initially excited wavenumbers remain.

Frequencies of gravity waves in the solar interior in the linear, non-dissipative, non-rotating case have been calculated previously (\citep{un89,cd02} and generally range from 100$\mu$Hz at low horizontal mode (large wavelength) to 500$\mu$Hz for higher horizontal mode (small wavelength) depending on the radial order.  Figure 7 shows the power spectra of gravity waves in the radiative interior in our partially linear simulation shortly after the simulation was started.  Significant power occurs at the base of the convection zone, up to the highest frequencies for the low modes and up to about 200$\mu$Hz for the high modes (figure 7a).  At this radius the fluid motion is fully nonlinear and the expected broadband nature of the convection is seen.  Note that at the convective-radiative boundary there is power in frequencies much higher than the typical convective turnover frequency of 2-10$\mu$Hz because descending plumes pummel the stable region at many locations slightly out of phase.  Power is not concentrated at the convective turnover frequency as is often assumed.

Moving into the stable layer, figure 7b shows the dispersion relation for gravity waves at 0.575 $R_{\odot}$.  The peak power in figures 7b and c is three orders of magnitude smaller than the peak power shown in figure 7a.  There is (significant) power at these smaller radii only out to horizontal mode 25, indicating the rapid decrease in energy due to radiative damping.  In figure 7b there is power in both low frequency propagating g-modes generated by the overlying convection and penetration (below 100$\mu$Hz) and high frequency (above 100$\mu$Hz) standing waves.  Ridges similar to those calculated by asymptotic analysis (Christensen-Dalsgaard 1986) for high radial order are apparent.  At any given height there are several radial orders contributing to the power at that horizontal mode and frequency, giving rise to the width of the ridges seen in this dispersion relation.  At radius 0.360 $R_{\odot}$ (figure 7c), it is seen that power at higher modes is diminished relative to figure 7b and that the lowest frequency/high wavenumbers have been completely damped (see section 4.2.2).  Again, there is significant energy in the propagating waves at frequencies much higher than the convective turnover time.  

\subsection{Nonlinear Gravity Waves}
Figure 8a shows a snapshot of the vorticity produced with our fully nonlinear model for the full computational domain (as in figure 3b).  Figure 8b shows the temperature perturbation just in the radiation zone to illustrate the salient features of the gravity waves produced; waves spiral toward the centre due to rotation and geometry.  While both small and large horizontal modes are produced at the interface only small mode numbers survive at the centre with substantial amplitude.  

Figure 9 shows the power spectra for our fully nonlinear case after the convection has reached a statistically steady state.  The peak amplitude in figure 9 is scaled the same as in figure 7.  There is little energy in the high frequency standing waves.  That is, there is little power at frequencies above 100$\mu$Hz for modes greater than 10, in figures 9b and c.  The lack of standing modes after the convection has reached steady state is seen in both linear and nonlinear calculations, indicating that the energy in these frequencies is radiatively damped and is not continually generated by the overlying convection.  

The appearance of the high frequency standing waves is a numerical artifact and we show them here purely as a proof that the model can reproduce the asymptotic limit (given a large enough perturbation).  The simulations are started with an initial perturbation to the convection zone.  This initial perturbation has much higher amplitude than steady state convection.  The large amplitude perturbation hits the stable region hard enough to excite all resonant frequencies.  Once the convection has reached a statistically ``steady'' state, high frequencies still exist in the convection, but at a lower amplitude.  These lower amplitudes are unable to drive the high frequency waves.  

The (dis)appearance of the high frequency standing g-modes was investigated in several different models.  We ran a series of models including: partially linear, fully nonlinear, increased rotation rate, decreased rotation rate, larger and smaller diffusion coefficients.  In general any ``large'' perturbation to the momentum equation, like increasing or decreasing the rotation rate by a large value, or turning off nonlinear terms gives rise to the high frequency standing modes; however, these modes disappear on a timescale of $\approx 2-5\times10^{6}$s.  It appears that statistically ``steady'' convection does not provide a large enough perturbation to continually generate these modes.

In the Sun the thermal diffusivity is $10^{5}$ times smaller than in our simulation and therefore, one would expect these high frequency waves to damp on the order of $10^{11}$s rather than the $10^{6}$s seen in our simulation.  However, this is still a relatively short time; and therefore these high frequency standing waves likely do not exist in the Sun.  In addition, it has been found in previous studies that two dimensional simulations produce larger perturbations to the underlying stable region than 3D simulations.  In 3D these modes would be excited with lower amplitude and hence, would be even less likely to exist.  In order for these modes to exist, they will have to be continually generated by something other than steady convection.

\subsubsection{Radiative Damping}

As waves propagate in a diffusive medium they are radiatively damped.  This damping leads to the transfer of their angular momentum, positive or negative, to the differential rotation. The radiative damping of waves depends on the frequency and wavenumber of the waves.  Low frequency and high wavenumber (small wavelength) waves are damped more quickly (over a shorter distance) than high frequency/low wavenumber waves.  Because thermal diffusion has more time and larger temperature gradients for the former.

In this geometry group velocity is directed inward and phase velocity is outward.  The inward propagation of wave packets, indicative of group velocity, can be seen in figure 10.  This figure shows the temperature perturbation as a function of radius for a given horizontal mode (in this case m=20 and 100, where the horizontal mode number m is defined as $k_{h}$r) for two sequential times, with the dotted line representing the earlier time and the solid line representing a later time ($10^{4}$s later).  It can be seen that wave packets move inward in time.  As wave packets move inward they are radiatively damped and therefore, their amplitudes decrease.  In figure 10, not only does the lower m case have higher amplitude overall, it is damped less in time relative to the higher m case.  In animations of these data a group velocity can be (crudely) measured and we find that at these radii (far away from the transition) this group velocity agrees well with that predicted by the linear dispersion relation.  While we are able to pick out wave packets and a group velocity for higher values of the horizontal mode m, we see only standing waves in lower values of m.  It is likely that there is a propagating wave component, however, it has significantly less amplitude than the standing waves and is therefore, hard to discern.  This has implications for energy in propagating waves (section 4.2.2).      

The radiative damping of waves can also be seen in figure 11.  In that figure we show the kinetic energy density in waves as a function of radius for three different frequencies and horizontal mode numbers.  The energy at the convective-radiative interface varies by no more than an order of magnitude across the three different frequencies and modes shown.  However, moving radially inward it is seen, as expected, that the lowest frequency waves are damped more rapidly with depth, with the 2$\mu$Hz, m=1 wave amplitude falling by nearly 6 orders of magnitude and the 10 and 20$\mu$Hz,m=1 waves dropping by only around 3 orders of magnitude.  Also as expected, higher values of horizontal mode numbers are damped more than m=1 and their energy in the lower frequencies is damped over a shorter distance.  For comparison, the dot-dashed line shows the analytic prediction for the damping of the m=1 wave \citep{ktz99}.  Using amplitudes for the kinetic energy at the convective-radiative interface from these simulations, we damp them with depth using the analytic prescription in \cite{ktz99}; there is a substantial difference, particularly for the lowest frequencies.  In figure 11 it is seen that (1) only the very low m waves make it to the centre with any appreciable energy, (2) all three frequencies are excited with remarkably similar amplitudes and (3) m=1 looks like a standing wave.  

\subsubsection{Energy in g-modes}

Gravity wave energy and flux is of great interest for mixing and angular momentum transport in the solar radiative region.  The time averaged kinetic energy density as a function of radius in the stable region is shown in figure 12. The large kinetic energy just below the convection zone is due to convective overshoot.  The local peak in energy density at the centre may be due to the focusing affect discussed by Press (1981).  We can also calculate the wave kinetic energy as a function of horizontal wavenumber ($k_{h}$) and frequency ($\it{f}$) at a particular radius: 

\begin{equation}
E(\it{\omega},k_{h}) = \frac{1}{2}\rho(v_{\theta}(\it{\omega},k_{h})^{2}+ v_{r}(\it{\omega},k_{h})^{2}).
\end{equation}
This can be compared with the energy calculated using (14) and (15) :

\begin{equation}
E(\it{\omega},k_{h}) = \frac{1}{2}\rho (\frac{N}{\omega})^{2}v_{r}(\it{\omega},k_{h})^{2}
\end{equation}
where $k_{h}=\frac{m}{r}$.  The (dis)agreement of these two equations gives us an estimate of the applicability of the linear dispersion relation at different radii, wavenumbers and frequencies.  Figure 13 shows these two relations at two radii and two horizontal modes as a function of frequency.  It is seen in that figure that the two formulations do not agree near the convective-radiative interface, with the linear dispersion relation underestimating the energy by 2+ orders of magnitude for both m=1 and m=10.  Well away from the convective-radiative boundary, the linear dispersion relation is a good approximation, particularly at m=10 where the two formulations match exactly (the dotted line is under the solid line).  It is expected that the linear dispersion relation would break down near the convective-radiative interface, given that overshooting plumes cause significant nonlinearity.  

Since the linear dispersion relation does not hold near the convective-radiative boundary it is difficult to justify the use of the group velocity calculated from that dispersion relation at those radii.  Therefore, in order to estimate the energy flux in waves at the base of the convection zone (defined as energy density in waves times the group velocity times the surface area), we use the group velocity at a deeper radius, where the linear dispersion relation is a good approximation.

The wave flux calculated this way is shown in figure 14.  The peak flux here (m=1,low frequency) is three orders of magnitude smaller than the peak wave flux predicted in \cite{ktz99} for l=1.  However, in our simulations this flux is distributed evenly in frequency for all (relatively low) wavemodes unlike the calculations by \cite{ktz99} where the flux drops off rapidly with frequency.  This broadband distribution of energy produced is seen in figure 11 and was found in the simulations by \cite{kir03}.   

These estimates should be considered rough at best, as the group velocity at radii close to the convective-radiative boundary is very uncertain.  The kinetic energy of the fluid, calculated using (17) is likely not just wave energy.  However, this energy, while not completely wave energy, still contributes to the angular momentum transport and mixing below the convection zone.  In addition, these estimates should be considered as upper limits on the wave flux for a couple of other reasons; (1) in two dimensions energy is transferred to the stable region more readily than in 3D, because of the alignment of convective cells and the ``flywheel'' motion that tends to occur in 2D convection because of the inverse cascade, (2) our thermal diffusivity is signficantly larger than the solar value (by a factor of $10^{5}$), causing the convective velocity to be larger than those expected in the Sun and therefore, the wave flux could also be larger and (3) most of the energy in low frequency/ low mode waves are in the standing wave component rather than the propagating wave (figure 11). 
 
\section{Conclusions}
We have presented self consistent calculations of gravity wave excitation by overlying convection and penetration.  These simulations produce power spectra of the gravity waves as a function of radius.  We find that both high frequency standing g-modes and low frequency propagating modes are excited depending on the amplitude of the perturbation.  In general the standing waves are not continually driven by the convection and only larger perturbations to the momentum equation produce these modes.  

Nonlinear effects need to be considered in the radiation zone for several reasons.  These effects broaden the frequency ridges in the dispersion relation, they allow for transfer of energy from high frequency to low frequency modes.  In addition, just beneath the convection zone, where plumes overshoot, nonlinear interactions increase the energy there by more than two orders of magnitude over what the linear dispersion relation would predict for energy in waves.  While this energy may not be completely wave energy it still contributes to the dynamics of the region just below the convection zone.   

The low frequency modes are driven by the overlying convection and penetration.  Typical frequencies of these waves can be ten times larger than the convective turnover frequency, which varies from 2-10$\mu$Hz.  We also find that the energy in these waves is distributed rather evenly in frequency.  The steep drop off in energy at larger frequencies is not observed.  

While these are self-consistent calculations of gravity waves in a realistic solar simulation, there are many shortcomings that still need to be addressed.  First is the three dimensional effect on the waves and probably more importantly on the nature of the convection which drives the waves.  In addition, any numerical simulation will be more laminar than the Sun.  Presumably, a more turbulent convection zone should provide a broader spectrum (in frequency and wavenumber) of waves.  What effect this will have on the spectrum and dissipation of the waves is unclear.  

\section*{Acknowledgments}
We thank J. Christensen-Dalsgaard and D. Gough for helpful discussions and guidance as well as an anonymous referee for insightful comments.  T.R. would like to thank the NPSC for a graduate student fellowship and the Institute of Astronomy, Cambridge University for a travel grant.  Support has also been provided by DOE SciDAC program DE-FC02-01ER41176, and the UCSC Institute of Geophysics and Planetary Physics.  Computing resources were provided by NAS at NASA Ames and by an NSF MRI grant AST-0079757.

\clearpage

\begin{figure}
\includegraphics{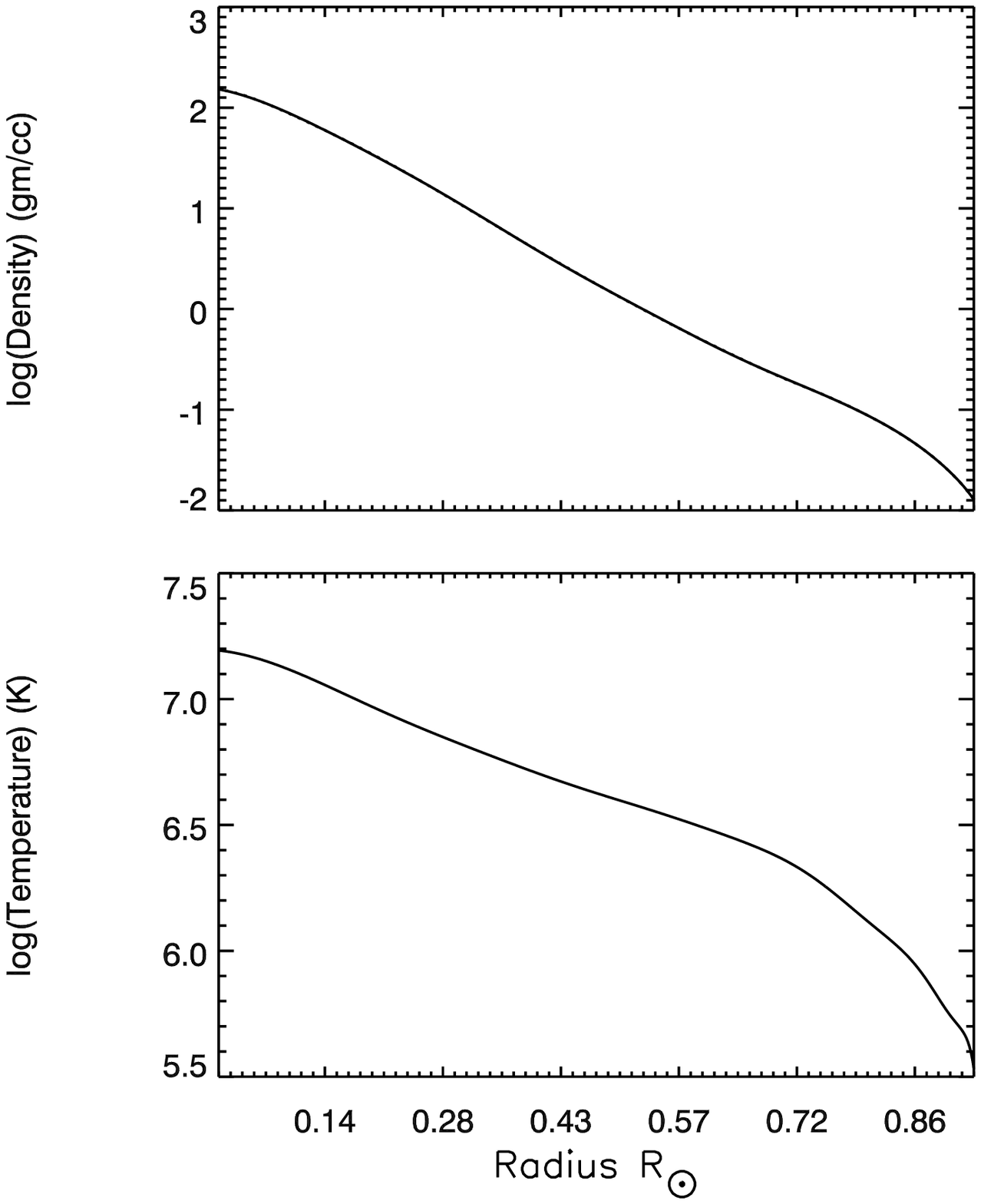}
\caption{Reference state density and temperature, taken from the 1D solar model.  Both fitted values (solid line) and actual values (dotted lines) are shown (dotted line is directly under the solid line).}
\end{figure}

\clearpage
\begin{figure}
\includegraphics{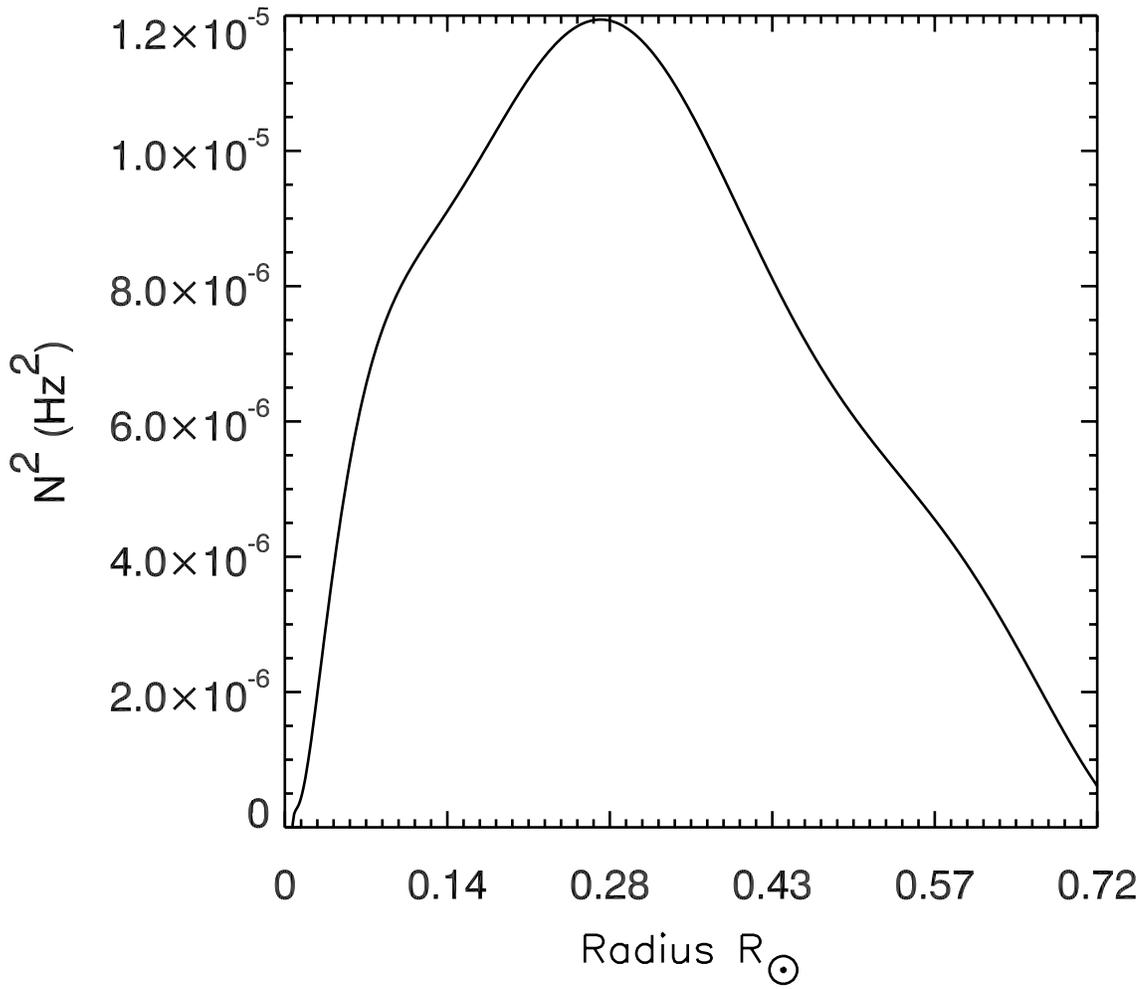}
\caption{Brunt-Vaisala frequency squared as a function of radius in the stable region.  Calculated using equation 16.}
\end{figure}

\clearpage

\begin{figure}
\includegraphics[width=5in]{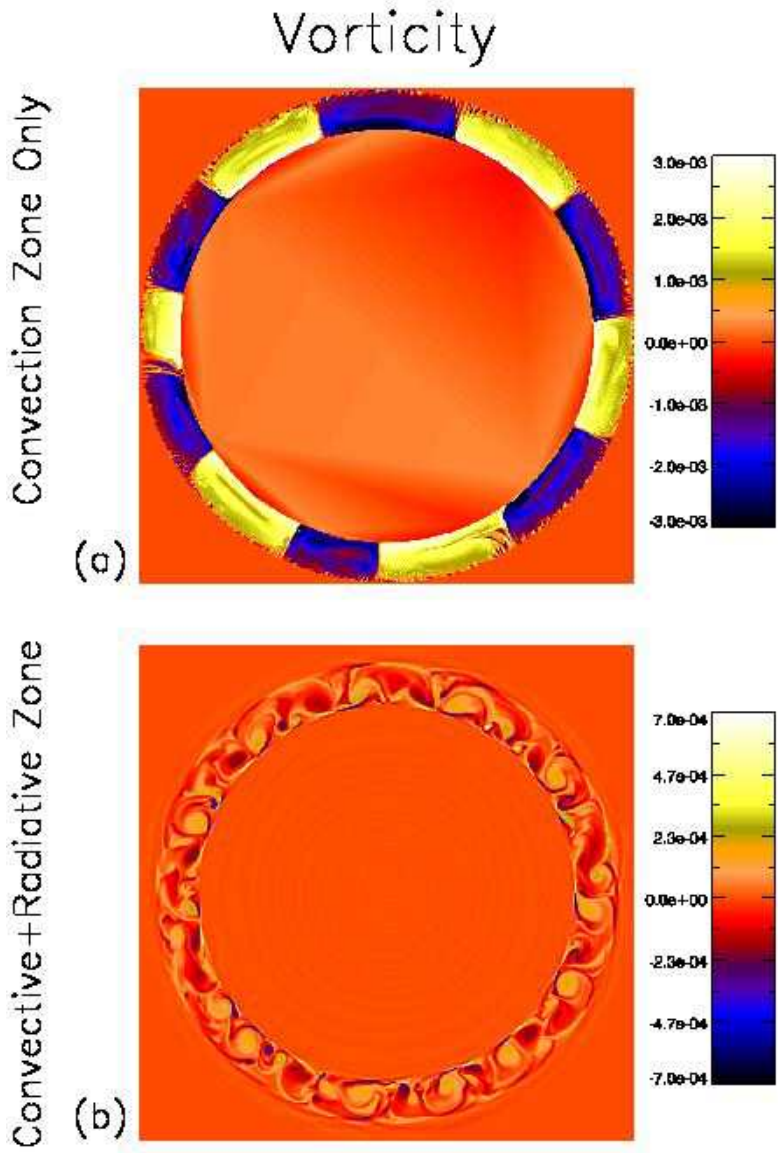}
\caption{Vorticity (in $\frac{rad}{s}$) for a convection model with impermeable boundaries (a) and for convection model with a stable layer below (b), blue represents negative vorticity while white/yellow represent positive vorticity.}
\end{figure}

\clearpage

\begin{figure}
\includegraphics{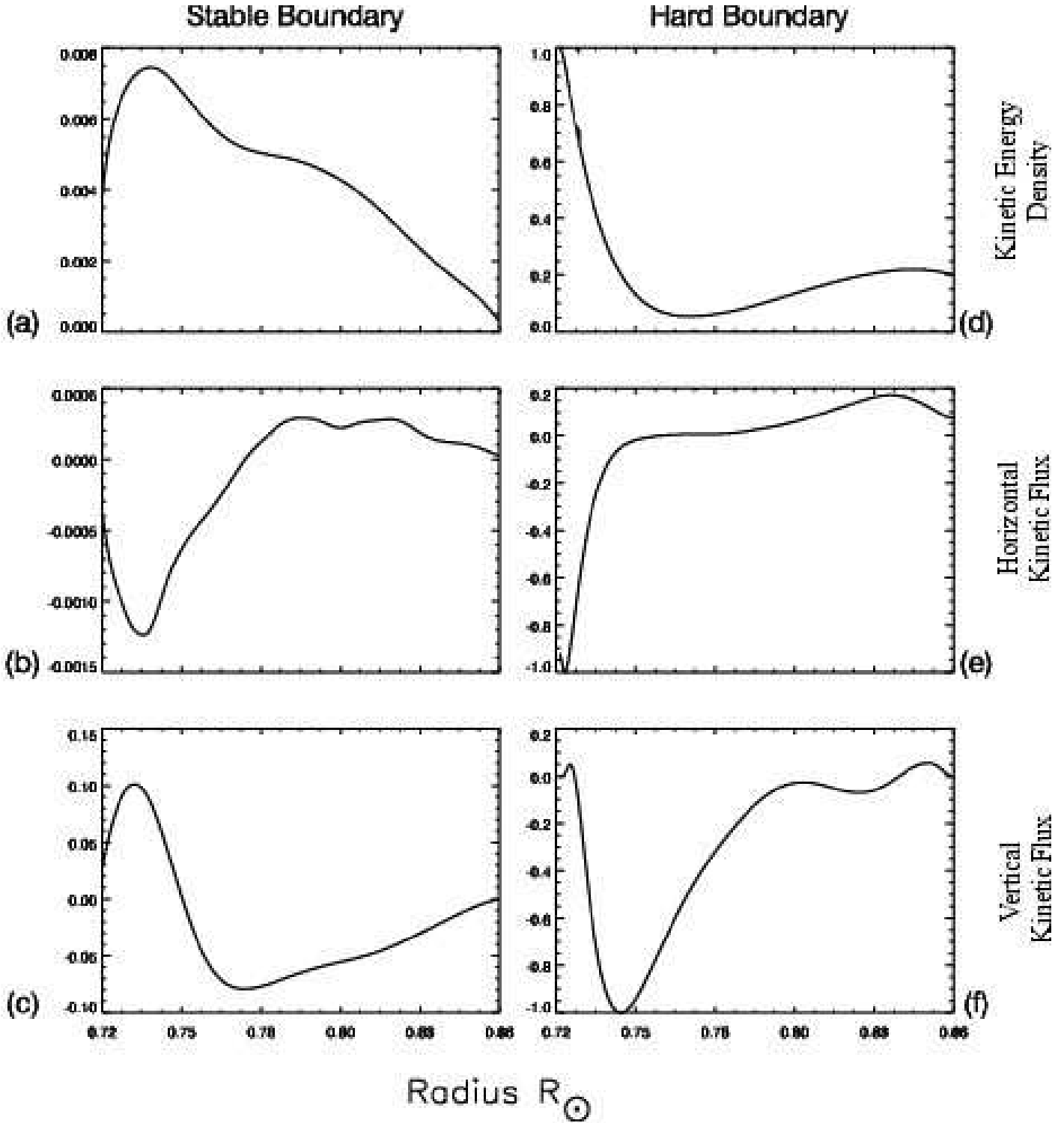}
\caption{Kinetic Energy Density ($\frac{ergs}{cm^{3}}$) (a) \& (d), Horizontal Kinetic Energy Flux ($\frac{ergs}{cm^{2}s}$) (b) \& (e) and Vertical Kinetic Energy Flux ($\frac{ergs}{cm^{2}s}$) (c) \& (f).  All plots are scaled to the peak value in the impermeable case.  The qualitative nature of the energy and energy fluxes are not vastly different, however, their amplitudes are, with the total energy density in the convective region of the impermeable case 2-3 orders of magnitude larger than in the model bounded by a stable layer.  A similar difference in magnitude is seen in the horizontal flux, while the vertical flux is closer in the two models.} 
\end{figure}

\clearpage

\begin{figure}
\includegraphics{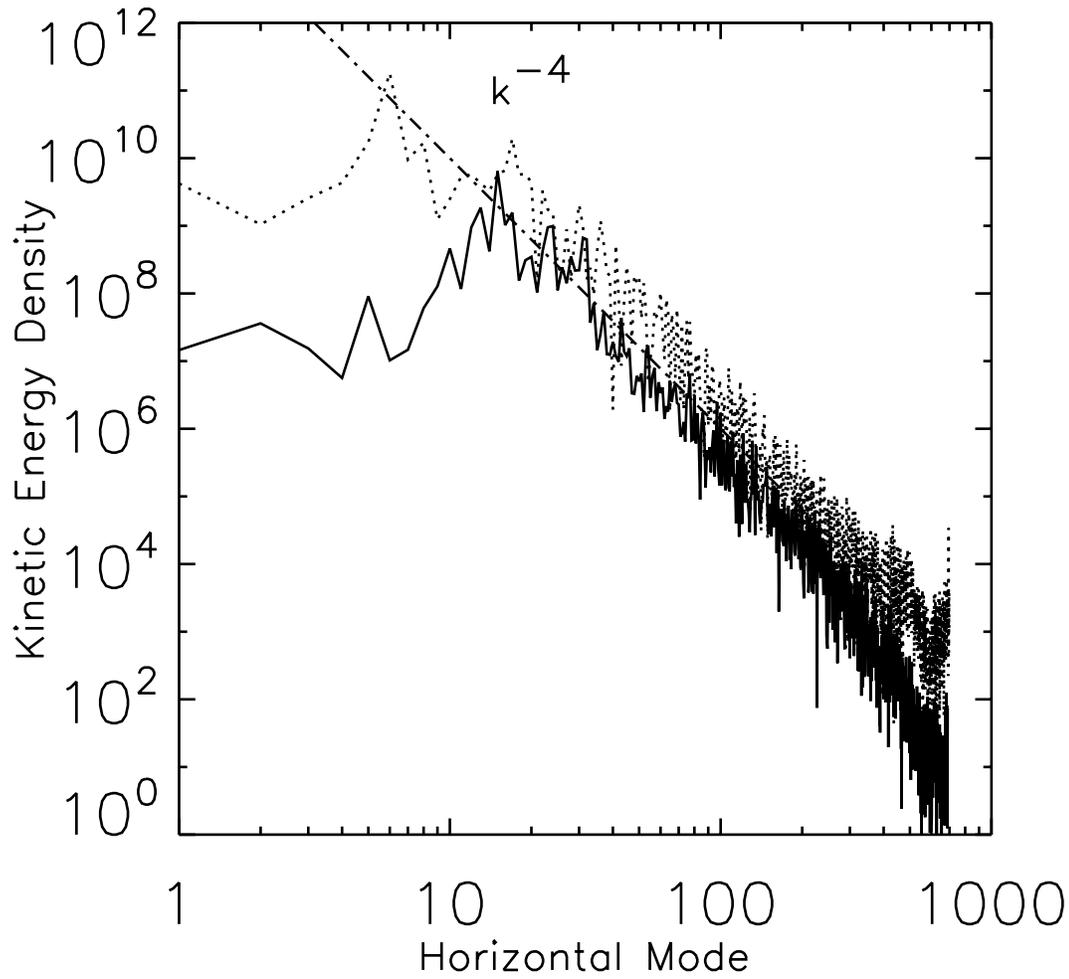}
\caption{Kinetic energy spectra for the model with an impermeable lower boundary (dotted line) and for the model with a radiative region at the bottom boundary (solid line).  The energy is greater at all horizontal wavenumbers in the impermeable model.  In addition, no dissipation region is seen in the impermeable model, indicating buildup of energy at the smallest scales in that case.}
\end{figure}

\clearpage

\begin{figure}
\includegraphics[width=5in]{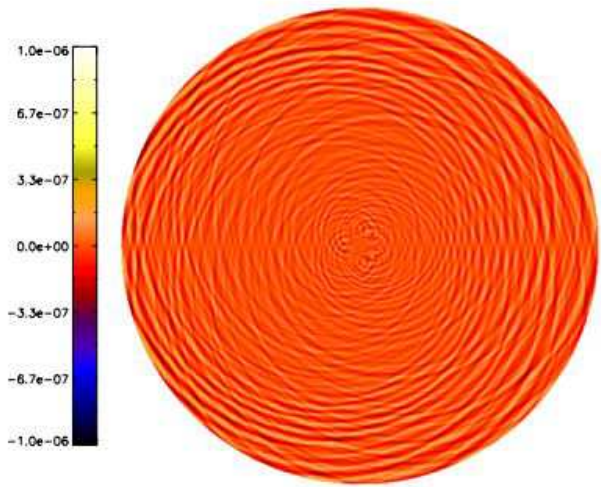}
\caption{A snapshot of gravity waves in the radiative interior in a completely linear model. Vorticity ($\frac{rad}{s}$) is shown.  This pattern is similar to that calculated from ray-tracing.}
\end{figure}

\clearpage

\clearpage
\begin{figure}
\includegraphics[width=5in]{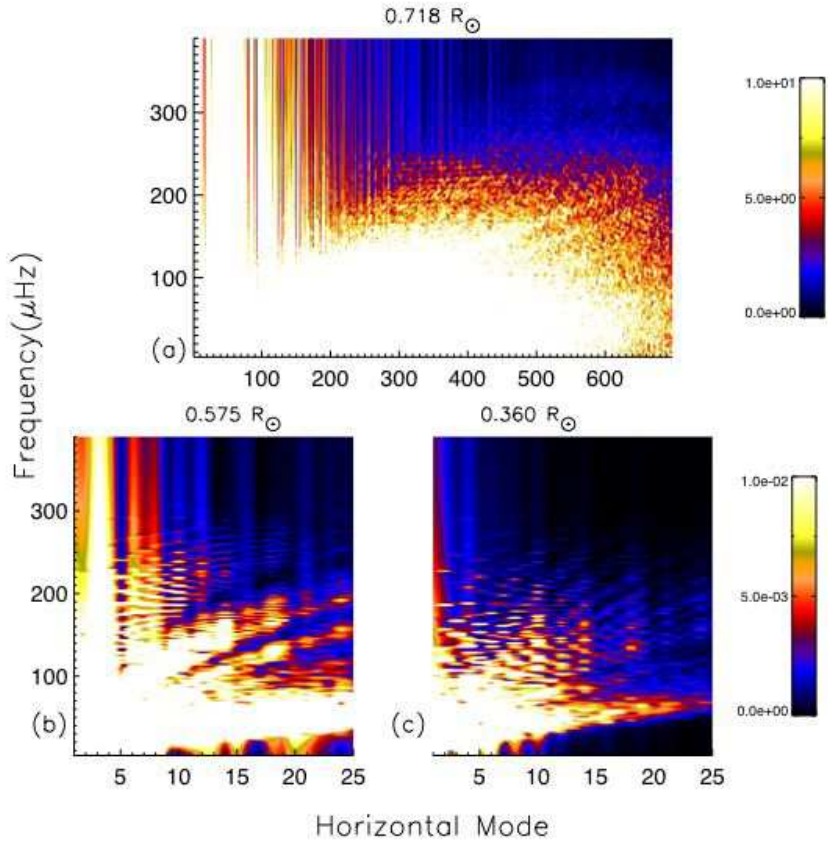}
\caption{Power spectra ($\frac{cm^{2}}{s^{2}}$) of gravity waves at the base of the convection zone (a) and deeper into the radiative interior (b)/(c) for our partially linear model.  White represents large energy, blue represents low energy.  Peak energy in (a) is three orders of magnitude larger than in (b)/(c).  Energy is broadband in frequency and horizontal wavenumber near the base of the convection zone.  Moving deeper into the radiative interior ridges from standing g-modes are seen at high frequencies, while low frequencies are occupied by propagating waves from the overlying convection.  These spectra were taken shortly after the simulation was begun.}
\end{figure}

\clearpage
\begin{figure}
\includegraphics[width=5in]{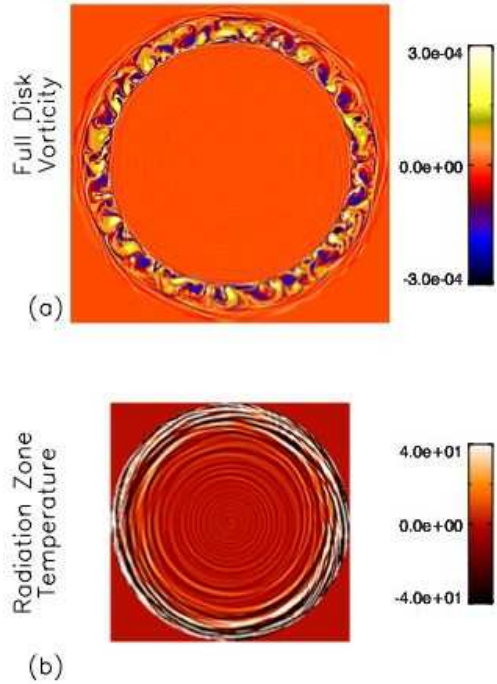}
\caption{Snapshots of our fully nonlinear model.  In (a) we show the vorticity ($\frac{rad}{s}$) in the full computational domain.  In (b) we show the temperature perturbation (white represents positive-hotter than reference state temperature, while black represents negative-colder than reference state temperature) in the radiative region only to emphasize the gravity waves.  It is seen there that the amplitude of the waves falls off quickly and that while many scales of g-modes are produced at the interface, only long wavelength waves survive in the center.  This gravity wave pattern is significantly different than that shown in figure 6.}
\end{figure}

\clearpage
\begin{figure}
\includegraphics[width=5in]{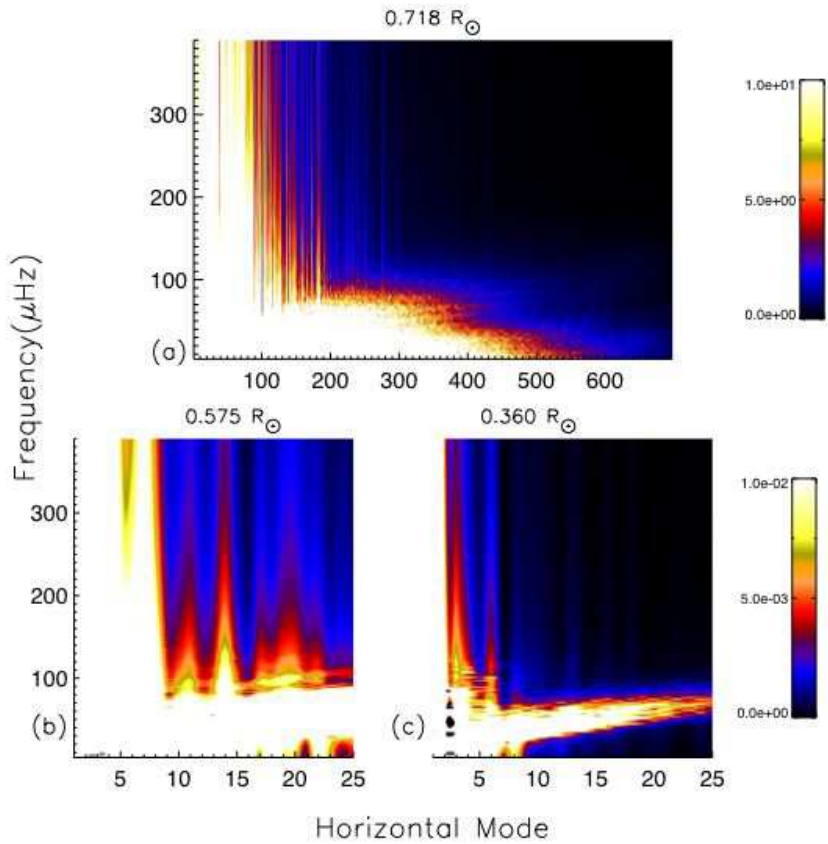}
\caption{Energy spectra, similar to figure 8 but for our fully nonlinear model and after the convection has reached steady state.  Within the radiation zone (figures b and c) the peak energy is three orders of magnitude larger than in the linear case; the nonlinear case transfers energy more efficiently to the radiative region.  However, the high frequency standing g-modes are gone.}
\end{figure}

\clearpage
\begin{figure}
\includegraphics{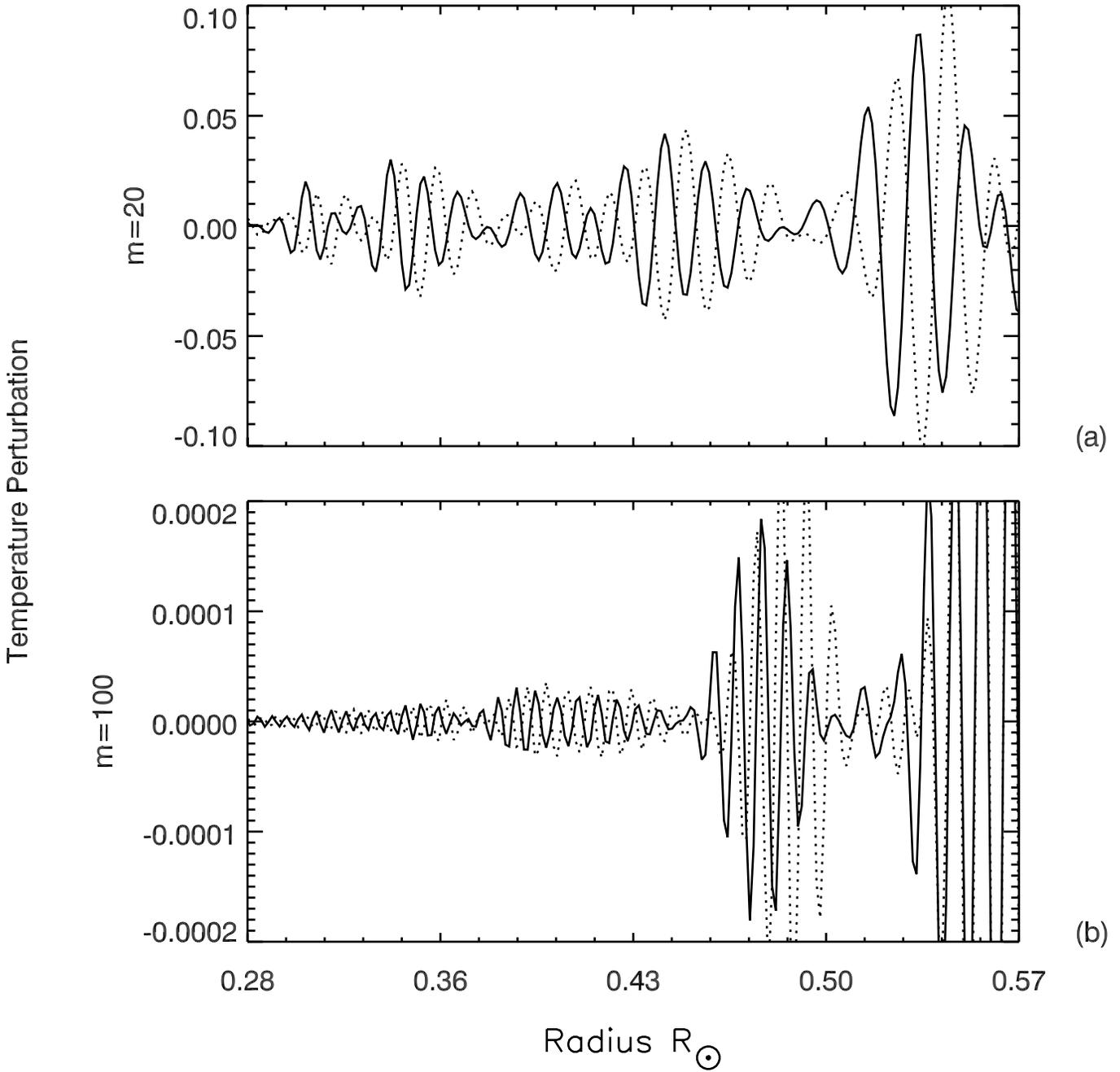}
\caption{Temperature perturbation within the radiation zone at two different times for horizontal modes m=20 and 100 are shown.  The dotted line represents an earlier time, while the solid line represents a later time, illustrating the inward movement of wave packets and group velocity. The decrease in amplitudes of the wave packets shows the radiative damping while moving radially inward.}
\end{figure}

\clearpage
\begin{figure}
\includegraphics{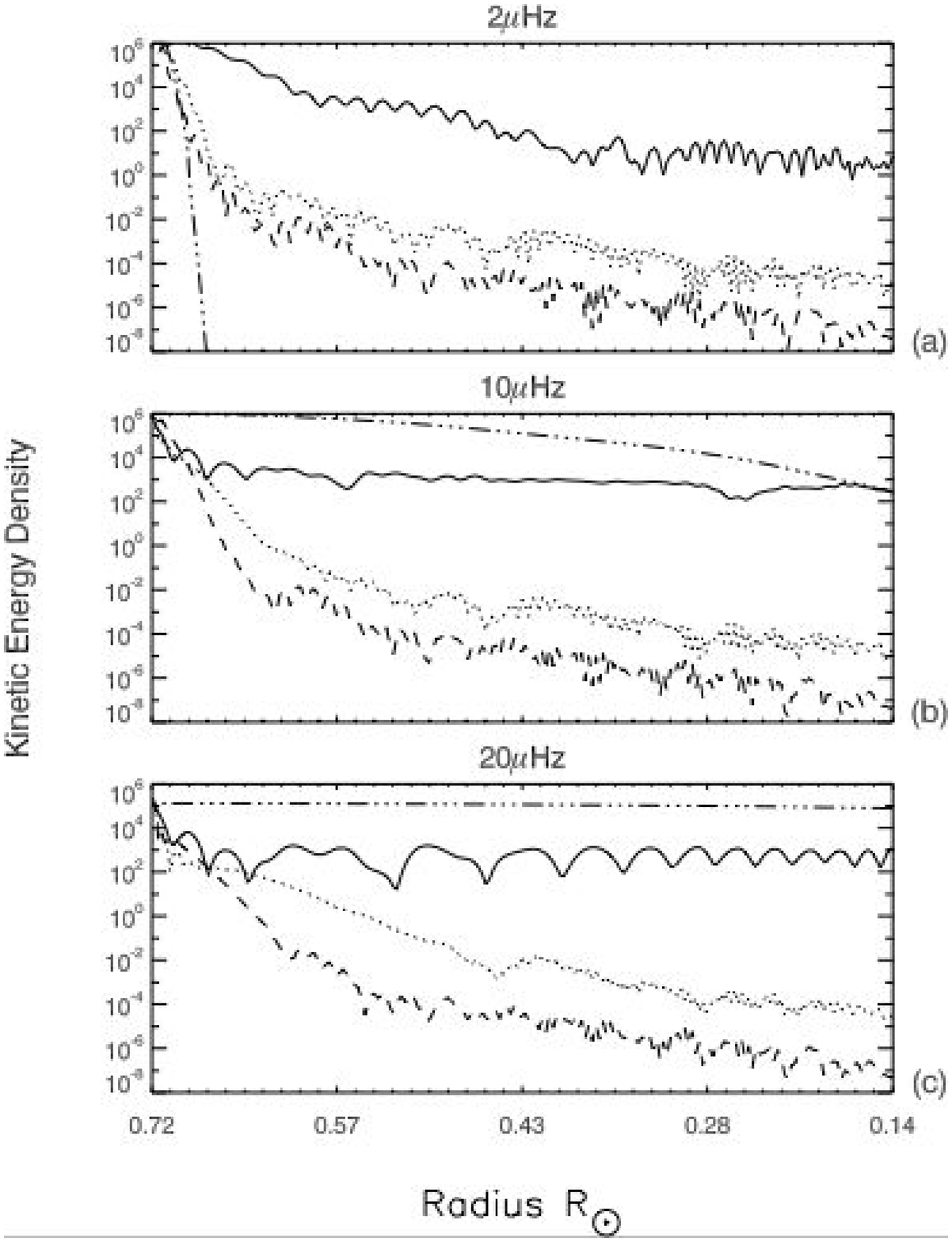}
\caption{Energy density ($\frac{gm}{cm s^{2}}$) as a function of radius for three frequencies: 2$\mu$Hz (a), 10$\mu$Hz (b) and 20$\mu$Hz (c) and three different horizontal modes: m=1 (solid line), m=10 (dotted line) and m=20 (dashed line) and for an analytic prediction of m=1 (dot-dashed line).  For m=1, 2$\mu$Hz waves are damped significantly more than their 10 and 20 $\mu$Hz counterparts.  Higher values of m are damped more rapidly with depth.  Only high frequency, low wavenumber waves make it to the center with any appreciable energy.  Note that all three frequencies are excited with similar amplitudes.  Note m=1 looks like a standing wave for at least 20 $\mu$Hz, although it could be argued for all m=1 waves shown.}

\end{figure}

\clearpage
\begin{figure}
\includegraphics{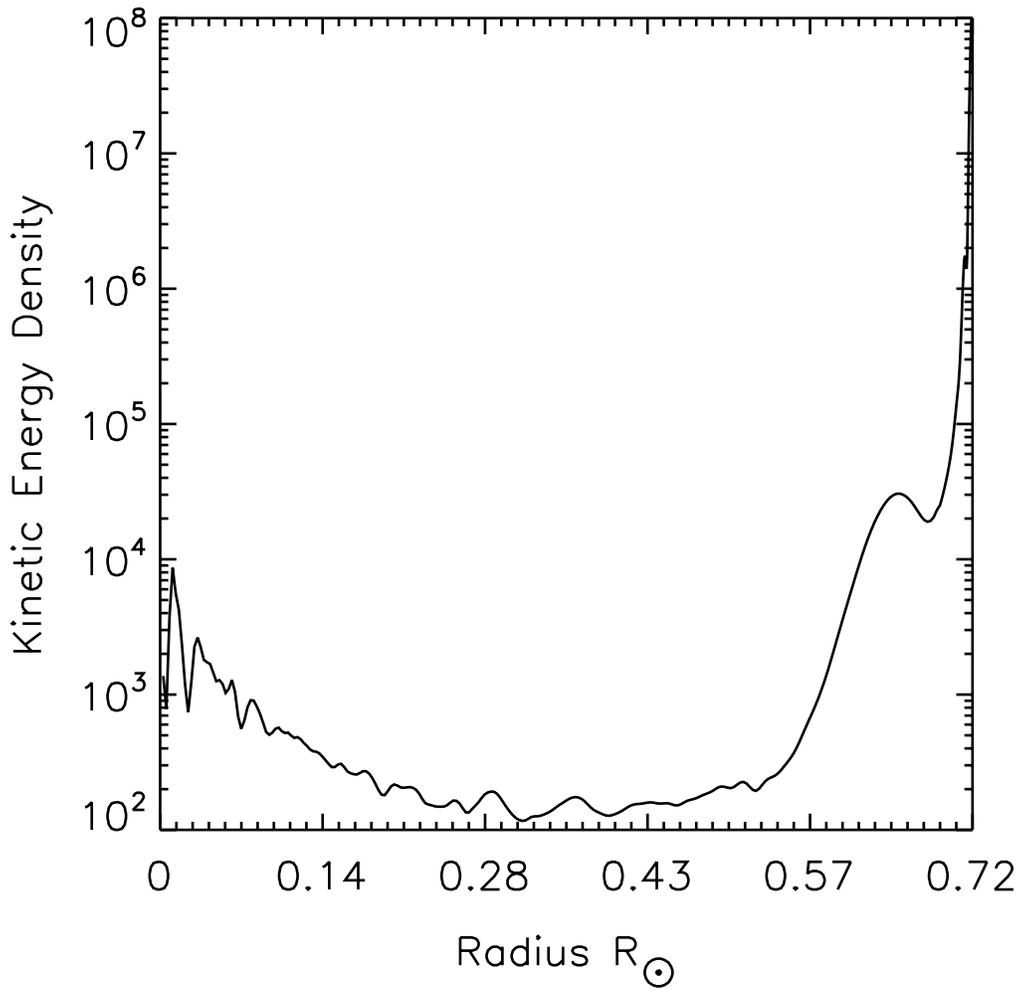}
\caption{Time averaged, real space, kinetic energy density ($\frac{gm}{cm s^{2}}$) as a function of radius within the radiative interior.  A rapid decrease in energy moving radially inward is seen immediately, but with some rise in energy closer to the center. }
\end{figure}

\clearpage
\begin{figure}
\includegraphics{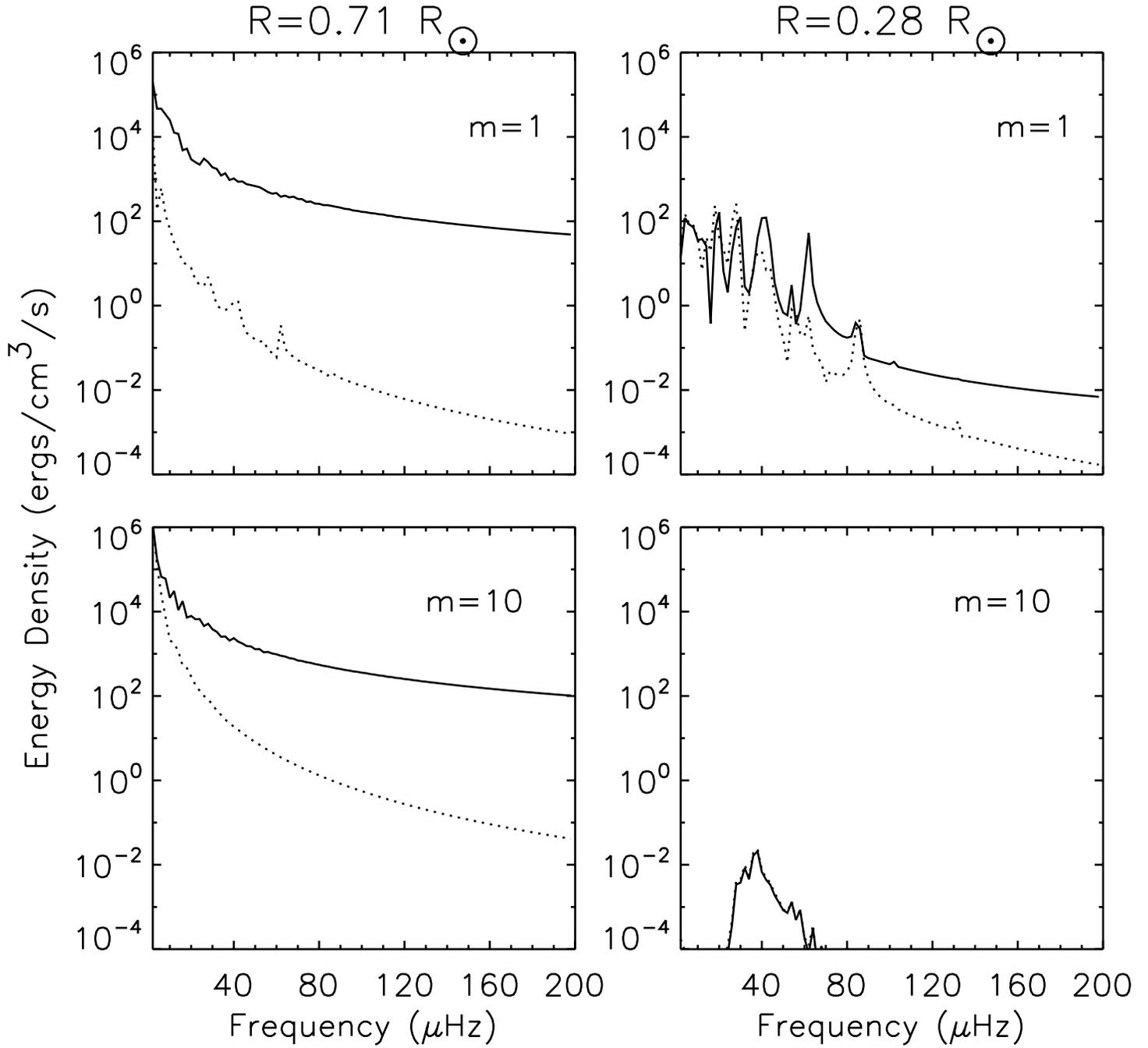}
\caption{Kinetic Energy density as a function of frequency, calculated as $\rho(v_{\theta}^2+v_{r}^2)$ solid line and as $\rho(\frac{N}{\omega})^{2}v_{r}^{2} $ (estimate from the linear dispersion relation) dotted line.  Near the convection zone the linear dispersion relation significantly underestimates the energy density for both modes shown.  However, deeper in the interior it is a better approximation.}
\end{figure}

\clearpage
\begin{figure}
\includegraphics{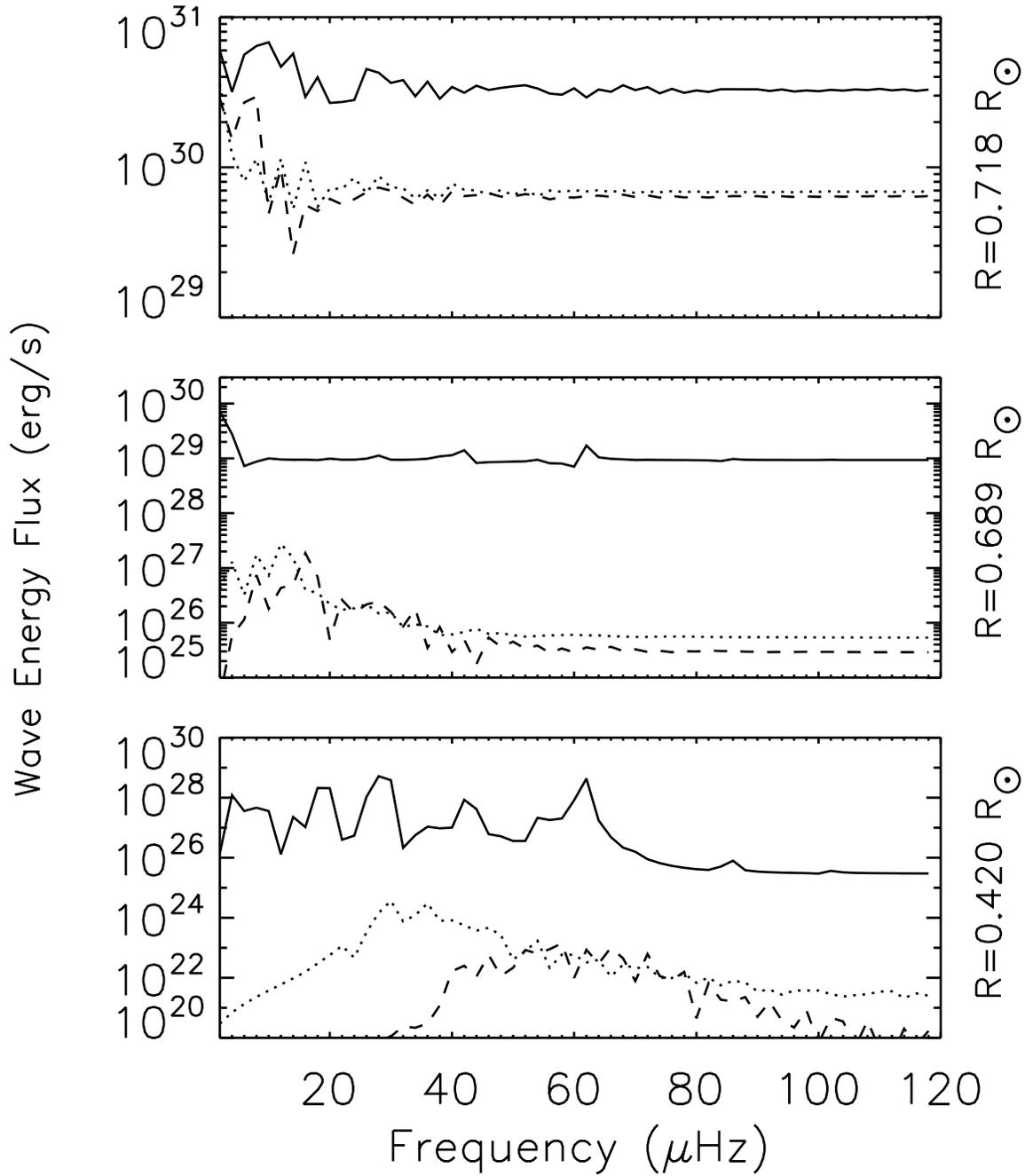}
\caption{Wave energy flux as a function of frequency for m=1 solid line, m=10 dotted line and m=20 dashed line, for three different radii within the radiation zone.  Wave energy flux is distributed rather evenly in frequency, particularly for m=1, unlike analytic estimates.}
\end{figure}

\label{lastpage}
\end{document}